\documentclass[]{aa}
\usepackage{graphicx}
\usepackage{txfonts}
\begin{document}
\title{BVR photometry of the resolved dwarf galaxy Ho\,IX
\thanks{
This work is a part of a joined project between the Astronomical
Institute of the Ruhr-University in Bochum and the Institute of
Astronomy of the Bulgarian Academy of Sciences for the study of 
nearby dwarf galaxies.}
}

\author{
Tsvetan B. Georgiev\inst{1}
\and 
Dominik J. Bomans\inst{3}
}

\institute{
Institute of Astronomy, Bulgarian Academy of Sciences and Isaak Newton
Institute of Chile, Bulgarian Branch, 
72 Tsarigradsko Chausse Blvd.  1784 Sofia, Bulgaria\\
\email{tsgeorg@astro.bas.bg}
\and 
Astronomisches Institut der Ruhr-Universit\"at Bochum, Universit\"atsstr. 
150, 44780 Bochum, Germany\\
\email{bomans@astro.ruhr-uni-bochum.de}
}

\offprints{Dominik J. Bomans}
\date{Received / Accepted}

\abstract{
We present BVR CCD photometry down to limiting magnitude B$=23.5$ mag 
for 232 starlike objects  
and 11 diffuse objects in a $5\farcm4 \times 5\farcm4$ field of Ho\,IX.  
The galaxy is a 
gas-rich irregular dwarf galaxy possibly very close to M\,81,  which 
makes it especially interesting in the context of the evolution of satellite
galaxies and the accretion of dwarf galaxies.  Investigations of Ho\,IX 
were hampered by relatively large contradictions in the magnitude scale 
between earlier studies.  With our new photometry we resolved these 
discrepancies.
The color magnitude diagram (CMD) of Ho\,IX is fairly typical of a 
star-forming dwarf irregular, consistent with earlier results.  Distance 
estimates from our new CMD are consistent with Ho\,IX being very close to 
M\,81 and therefore being a definite member of the M\,81 group, apparently
in very close physical proximity to M\,81.  
}
\titlerunning{Ho\,IX}
\authorrunning{Georgiev \& Bomans}
\maketitle
\section{Introduction}
\begin{figure}
\resizebox{\hsize}{!}{\includegraphics{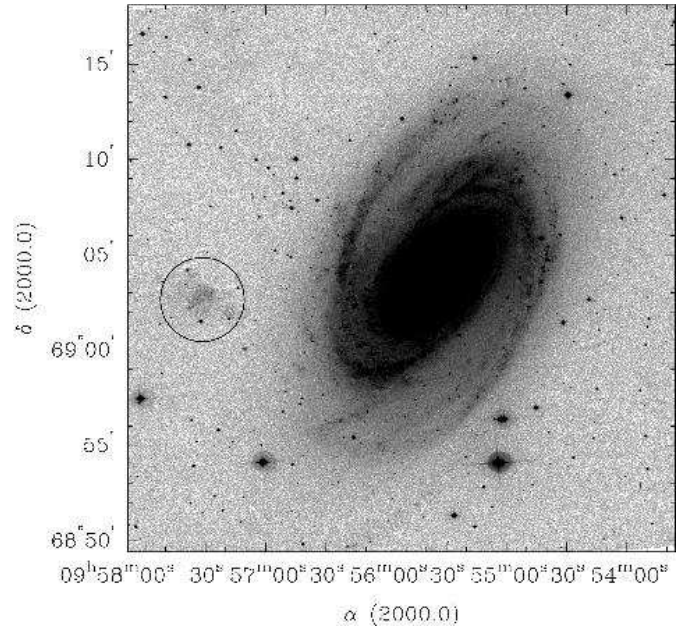}}
\caption{30\arcmin\ by 30\arcmin\ region around M\,81 from a DSS2 blue plate 
scan.  The dwarf galaxy Ho\,IX is visible as irregular low surface 
brightness region east of M\,81 and marked with a circle at its 
position.}
\label{fig1}
\end{figure}

The dwarf irregular galaxy Ho\,IX = DDO\,66 = UGC\,5336 = PGC\,28757 is 
a probable close 
companion of the nearby spiral galaxy M\,81. It has an apparent size of about 
2.6\arcmin\ by 2.2\arcmin, a total magnitude of B\,=\,14.41\,mag, integrated 
colors of B-V\,=\,0.2, U-B\,=\,$-0.4$, and a foreground extinction of 
A$_B = 0.32$ \cite{Paturel96}. 
Ho\,IX is a possibly gas-rich ($HI \sim -12.3$ mag) 
and low-luminosity (M$_B \sim -13.5$ mag) irregular galaxy with few
prominent young stellar associations and low overall surface brightness 
($\mu_B = 24.2$) \cite{Hopp87}. 
The projected location of Ho\,IX on the sky is very near to M\,81 
(see Fig.\,\ref{fig1}) and coincides with one of the 
neutral hydrogen clouds floating around M\,81 \cite{Yun94}. It is 
therefore very probable, that Ho\,IX has a very small physical distance from 
M\,81.  The complicated H\,I structure near M\,81 makes any association 
of H\,I to Ho\,IX quite uncertain.  Therefore the radial velocity (given 
as $46 \pm 6$\,km\,s$^{-1}$ in the NED database), and a 
distance based on the H\,I velocity of the H\,I detection coinciding 
with Ho\,IX is at least somewhat doubtful.

The first deep study of the stellar content of Ho\,IX, consisting of 41 
preliminary selected blue and red stars, was made by \cite{Sandage84q} on B 
and V photographs taken with the Palomar 5m telescope. The \cite{Sandage84q}  
sample was enlarged and measured on plates of the SAO 6m telescope by 
\cite{Georgiev91bq}. The photometric estimation of the distance modulus, 
based on the brightest blue and red stars, results in $(m-M)_0 = 27.7$ mag, 
which is practically equal to the distance modulus of M\,81 from Cepheid data 
of $(m-M)_o = 27.8$ mag, corresponding to $D = 3.6$\,Mpc \cite{Freedman94}. 

The stellar populations of Ho\,IX were studied using CCD detectors 
several times at the beginning of the `CCD era'. About 340 stars are measured 
in B, R, I by \cite{Hopp87q} using the Calar Alto 3.5\,m telescope and about 
250 in V, I using the 3.6\,m CFH Telescope by \cite{Davidge89q}. 
\cite{Hopp87q} investigated the distribution of the brightest 
blue and red stars and the color index of the unresolved background in Ho\
IX. They also found candidates for OB associations. The most remarkable of 
them is situated to the southeast from the center of Ho\,IX. 
The evolved stellar content of Ho\,IX was studied by \cite{Davidge89q}. They 
found that the number of 
evolved stars increases markedly at magnitudes fainter than I $= 21.5$\,mag 
and suggested this to be a consequence of a decrease in the rate of star 
formation roughly 50\,Myr ago. They also found that the luminosity function 
of the asymptotic giant branch stars in Ho\,IX is in good agreement with 
that measured in the LMC, suggesting that the intermediate-age 
star-formation history of Ho\,IX is similar to that of the LMC. 
   
The UIT images of Ho\,IX at wavelengths 249\,nm 
and 152\,nm, reported by \cite{Hill93q}, show no prominent young OB 
associations like those seen in the spiral arms of M\,81. They estimated  
that 60\% to 70\% of the V-flux of the galaxy is from an old population with
an age of about 10\,Gyr. The remainder of the stars formed about 20-200\,Myr 
ago. Individual stars measured in additional B and V frames appear to be 
evolved stars of mass about 12\,$M_0$ and an age of about 20\,Myr, belonging 
to the youngest population in the galaxy Ho\,IX. 
   
However, there are serious uncertainties about the magnitude scale in 
the field of Ho\,IX. In an attempt to improve the distance determinations 
to M\,81 and Ho\,IX, 
\cite{Metcalfe91q} (hereafter MS91) made B, V, R CCD observations of 
the sample of \cite{Sandage84q} using the 2.5\,m Isaac Newton Telescope. 
They found significant offsets 
between the magnitude zeropoints of the published data and the 
zeropoints of their photometry. 
Generally, the correct zeropoint transfer seems to be a very serious 
problem in the photometry of resolved galaxies, see also 
\cite{Hopp95q} and \cite{Aparicio97q}. 
   
The goals of the present work are (i) to survey the field of Ho\,IX 
with an independent transfer of the magnitude zero point, 
(ii) uses the new photometry to derive an improved distance to Ho\,IX, and 
(iii) construct photometric diagrams and luminosity functions of the blue 
stars to test the evolutionary status of the stellar content of 
Ho\,IX and the possible effects of its proximity to M\,81. 

\section{Observations and reductions}
The galaxy Ho\,IX has been observed with the 2m telescope of 
the Rozhen National Astronomical Observatory, equipped with a 
$1024 \times 1024$ pixel Photometrics CCD camera on Feb 2, 2000. 
The resulting 
readout noise is 5.1\,e-. The scale is 0.32\arcsec/pix, the frame size is 
5.4\arcmin\ by 5.4\arcmin. The air mass during the observations was 
1.12 and the seeing was 1.4\arcsec. The exposure times in B, V and R bands 
were $2 \times 20$, $2 \times 10$, and $2 \times 5$\,min, respectively. 
The exposure times were selected to achieve approximately equal depth in the 
different bands for stars with intermediate color indices. 
   
The reduction was performed in the same way as described in the papers of 
\cite{Makarova97q}, \cite{Georgiev97bq}, and \cite{Georgiev99q} using  
the Rozhen software \cite{Georgiev95b}. This software is an extension of 
the package PCVISTA of \cite{Treffers89q}. The    
reduction of the frames consists of the following steps: cleaning of cosmics, 
rebinning to a common pixel grid, smoothing the noise and selecting the 
objects on the basis of 
their magnitudes, sizes and shapes. Detailed information about the methods, 
especially of those methods used for photometry, is given in the Appendix A. 

The magnitudes were transferred from photometric standard stars in the open 
cluster NGC\,7790 \cite{Christian85} using an aperture diameter of 33 pix 
($\sim 10\arcsec$). The observations of Ho\,IX and NGC\,7790 are made during 
a stable photometric night at almost equal zenith distances. 
The color equations, without accounting for the air mass are: 
\begin{eqnarray}
B-b = 2.029 + 0.066(b-v)   (\pm 0.056) \\
V-v = 2.562 + 0.008(v-r)   (\pm 0.034) \\
R-r = 2.751 + 0.027(v-r)   (\pm 0.036) 
\end{eqnarray}

The aperture size of the presented photometry in Ho\,IX is 11 pix or $\sim
3.7\arcsec$. The aperture corrections are between 0.6 and 0.8\,mag. On the 
basis of our previous work we consider our total error (including 
the photometric error) to be $\sim 0.1$ 
mag in the region of V = 21\,mag and 0.2-0.3\,mag in the region of V = 
23\,mag. 
   
\begin{figure}
\resizebox{\hsize}{!}{\includegraphics{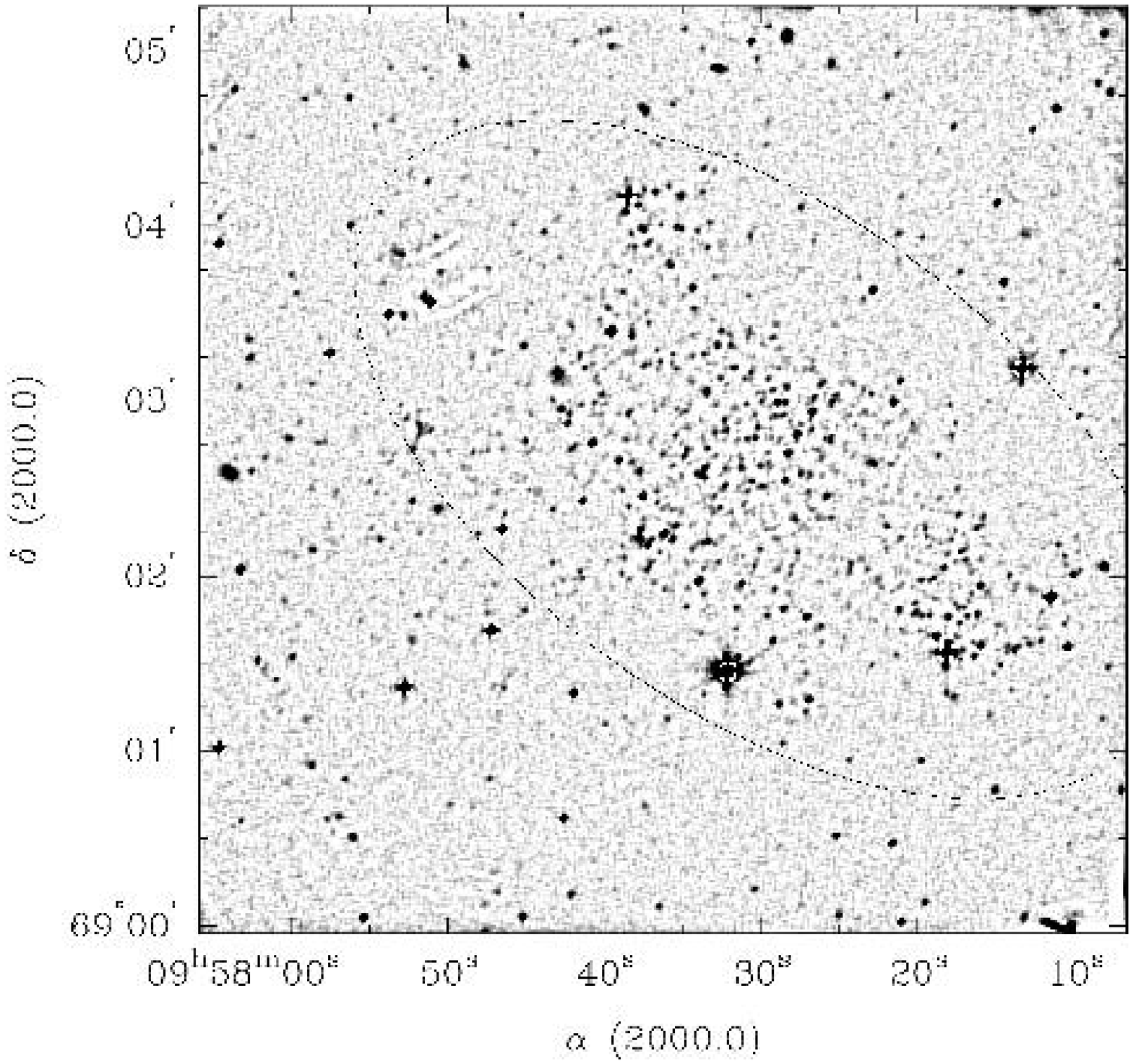}}
\caption{ 
Processed image of the observed field of Ho\,IX with size 5.4\arcmin\ 
by 5.4\arcmin. The coordinates were generated using a DSS2 image centered 
on Ho\,IX and the astrometry routines in IRAF.
}
\label{fig2}
\end{figure}

The map of the frame of Ho\,IX is given in Fig\,\ref{fig2}. It is made from 
the sum of the residual frames in the B, V and R bands, derived from the 
original frames by median filtering with a round window of 33\,pix (10\arcsec) 
size. The limiting magnitude of the map is about 24\,mag. The center of 
the frame is shifted to the west for better observation of the faint 
eastern periphery of Ho\,IX. Numerous faint stars with V $\sim$ 24\,mag are 
suspected there, but the depth of our frames is not sufficient for reliable 
photometry at these faint light levels.

\begin{figure}
\resizebox{\hsize}{!}{\includegraphics{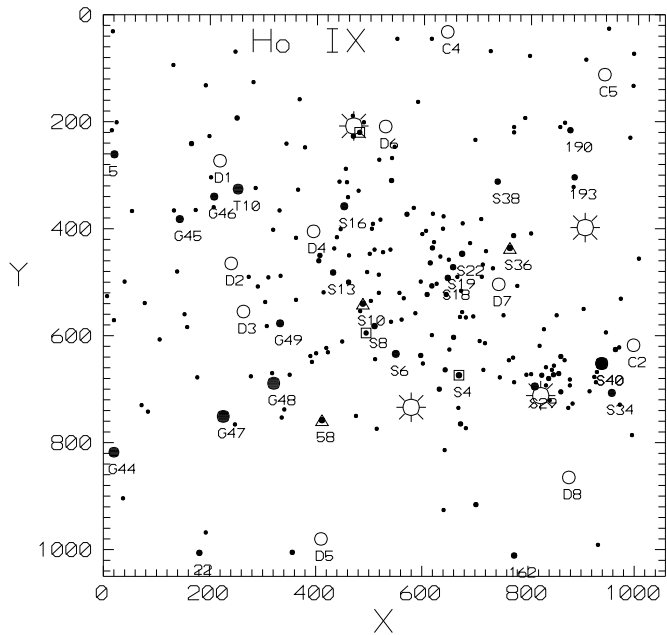}}
\caption{ 
Identification chart of the measured objects in the field of Ho\,IX. 
The XY coordinates in pixels correspond to the XY coordinates given 
in Table\,\ref{tab1}.  North is up and east is left on the diagram, as in 
Fig.\,\ref{fig2}.  The star-like objects are represented as filled dots, 
the extended objects as open circles.  The four brightest stars in the field, 
which where not measured by us, are marked with suns for orientation.
The brighter stars are also noted with their number from Table\,\ref{tab1}. 
The brightest blue and red supergiants used for the distance estimation 
are marked with additional boxes or triangles. 
}
\label{fig3}
\end{figure}

\begin{table*}
\begin{tabular}{|lcccccl||lcccccl|}
\hline
 ID &    X &  Y  &    V   &  B-V  &  V-R  & note &  ID &    X &  Y  &    V   &  B-V  &  V-R  & note\\
\hline
 S2 &  633 & 700 &  21.53 &  0.15 &  0.13 & s,d  &  30 &  252 & 193 &  21.65 &  0.49 &  0.32 & i\\
 S3 &  644 & 664 &  21.42 &  0.08 &  0.41 & s    &  56 &  408 & 450 &  21.95 &  0.11 &  0.18 &i\\
 S4 &  670 & 674 &  20.66 &  1.63 &  1.18 & s    &  73 &  471 & 227 &  21.53 &  0.07 &  0.53 &i\\
 S5 &  598 & 637 &  21.09 &  0.36 &  0.39 & s    &  77 &  483 & 220 &  21.64 &  1.65 &  1.12 &i\\
 S6 &  551 & 634 &  19.77 &  0.08 &  0.25 & s    &  80 &  495 & 595 &  21.13 &  1.34 &  0.92 &i,d\\
 S8 &  511 & 582 &  20.02 &  0.37 &  0.32 & s    & 112 &  610 & 523 &  21.63 &  0.94 &  0.60 &i\\
 S9 &  495 & 595 &  21.13 &  1.34 &  0.92 & s,d  & 113 &  616 & 464 &  21.68 &  0.78 &  0.55 &i\\
S10 &  489 & 540 &  20.35 &  0.38 &  0.33 & s,d  & 115 &  620 & 436 &  21.97 &  0.49 &  0.16 &i\\
S12 &  462 & 500 &  21.08 & -0.03 &  0.16 & s    & 116 &  619 & 507 &  21.52 &  0.43 & -0.07 &i\\
S13 &  433 & 482 &  20.69 &  1.49 &  0.84 & s    & 137 &  677 & 390 &  22.82 &  0.43 &  0.76 &i\\
S15 &  406 & 460 &  21.32 &  1.41 &  1.07 & s    & 138 &  673 & 765 &  21.09 &  1.50 &  1.72 &i\\
S16 &  454 & 358 &  19.13 &  0.00 &  0.39 & s    & 171 &  826 & 674 &  21.10 &  0.37 &  0.16 &i\\
S17 &  646 & 523 &  21.18 &  0.00 &  0.20 & s    & 175 &  839 & 680 &  21.62 &  1.03 &  0.51 &i\\
S18 &  649 & 492 &  20.72 &  0.06 & -0.00 & s    & 179 &  848 & 673 &  21.73 &  0.26 &  0.57 &i\\
S19 &  659 & 472 &  20.40 &  0.39 &  0.46 & s    & 182 &  858 & 671 &  21.39 &  0.21 &  0.29 &i\\
S22 &  676 & 447 &  20.52 &  0.30 &  0.41 & s    &   5 &   21 & 261 &  19.36 &  1.38 &  1.33 &o\\
S26 &  572 & 373 &  21.39 & -0.08 &  0.22 & s,d  &  19 &  166 & 241 &  21.41 &  1.16 &  1.49 &o\\
S27 &  543 & 310 &  21.07 &  0.08 &  0.22 & s    &  22 &  181 &1006 &  20.15 &  0.55 &  0.53 &o\\
S29 &  813 & 695 &  19.98 &  0.79 &  0.73 & s    &  47 &  356 &1005 &  21.02 &  0.51 &  0.63 &o\\
S30 &  862 & 705 &  21.05 & -0.18 &  0.33 & s    & 144 &  702 & 916 &  21.28 &  0.13 &  0.27 &o\\
S33 &  862 & 639 &  21.12 & -0.24 &  0.39 & s    & 162 &  774 &1011 &  20.99 &  1.32 &  1.38 &o\\
S34 &  958 & 707 &  19.50 &  1.39 &  1.10 & s    & 190 &  880 & 216 &  20.90 &  0.91 &  1.35 &o\\
S35 &  965 & 626 &  21.04 &  0.60 &  0.44 & s    & 193 &  888 & 304 &  20.32 &  0.66 &  0.57 &o\\
S36 &  766 & 436 &  20.54 &  0.27 &  0.28 & s    &  C2 &  999 & 618 &  20.49 &  1.04 &  0.75 &c\\
S37 &  773 & 413 &  21.37 &  0.02 &  0.42 & s    &  C4 &  649 &  32 &  19.28 &  0.57 &  0.50 &c\\
S38 &  743 & 312 &  20.43 &  1.18 &  0.88 & s    &  C5 &  945 & 112 &  20.47 &  0.70 &  0.56 &c\\
S40 &  939 & 652 &  17.45 &  0.82 &  0.67 & s    &  D1 &  220 & 273 &  21.14 &  0.06 &  0.34 &d\\
S41 &  483 & 220 &  21.64 &  1.65 &  1.12 & s    &  D2 &  241 & 465 &  21.46 &  1.27 &  1.06 &d\\
S44 &   20 & 818 &  18.56 &  1.30 &  1.26 & s    &  D3 &  264 & 555 &  21.03 &  0.08 &  0.25 &d\\
S45 &  144 & 382 &  19.41 &  1.22 &  1.50 & s    &  D4 &  396 & 405 &  20.68 &  1.07 &  1.08 &d\\
S46 &  209 & 340 &  19.76 &  1.42 &  1.48 & s    &  D5 &  410 & 980 &  21.41 &  0.61 &  0.90 &d\\
S48 &  321 & 689 &  17.73 &  1.27 &  0.92 & s    &  D6 &  532 & 209 &  21.07 &  0.90 &  1.01 &d\\
S49 &  333 & 577 &  19.66 &  1.29 &  1.36 & s    &  D7 &  745 & 504 &  21.03 &  0.09 &  0.39 &d\\
S50 &  412 & 758 &  20.37 & -0.04 &  0.12 & s    &  D8 &  877 & 865 &  21.27 &  0.77 &  1.12 &d\\
T10 &  254 & 326 &  18.20 &  0.38 &  0.45 & s    &     &      &     &        & &     &   \\
\hline
\end{tabular}
\caption{Data for the stars brighter than V $<$ 22 mag and for the diffuse 
objects in the observed field of Ho\,IX. The letters in the column `note'
have the following meaning: s = star measured by Sandage, i = inside the 
ellipse, o = outside the ellipse, c = globular cluster candidate, d = other 
diffuse objects. 
}
\label{tab1}
\end{table*}

\section{Resolving magnitude scale discrepancies}
MS91 made comparisons with the magnitudes of \cite{Sandage84q},
\cite{Hopp87q}, and \cite{Davidge89q}. All magnitude scales 
appear linear but with significant offsets: 
\begin{eqnarray}
V(MS) - V(D\&J)   = -0.36 \pm 0.23 \\
B(MS) - B(Sand.)  =  0.22 \pm 0.20 \\
V(MS) - V(Sand.)  =  0.06 \pm 0.27 \\
B(MS) - B(H\&S-L) = -1.01 \pm 0.36 \\
R(MS) - R(H\&S-L) = -1.09 \pm 0.38 
\end{eqnarray}

The large standard errors of the offsets may be due to the relatively big 
pixels of the CCD observations, e.g. in MS91 the pixel scales of 
the two CCDs used are 0.54\arcsec/pix and 0.74\arcsec/pix and the photometric 
aperture is 3x3 pix. 
   
Due to the nonlinear photographic extension of the magnitude 
scale down to 23\,mag, based on standards of 16 mag, the faintest stars 
in the work of \cite{Georgiev91bq} are fainter than those 
in \cite{Sandage84q} by 
about 0.4\,mag in V and 0.6\,mag in B band. For this reason the red 
supergiant candidates in Ho\,IX in \cite{Georgiev91bq} seem redder 
than in \cite{Sandage84q} by about 0.2\,mag. 

\begin{figure}
\resizebox{\hsize}{!}{\includegraphics{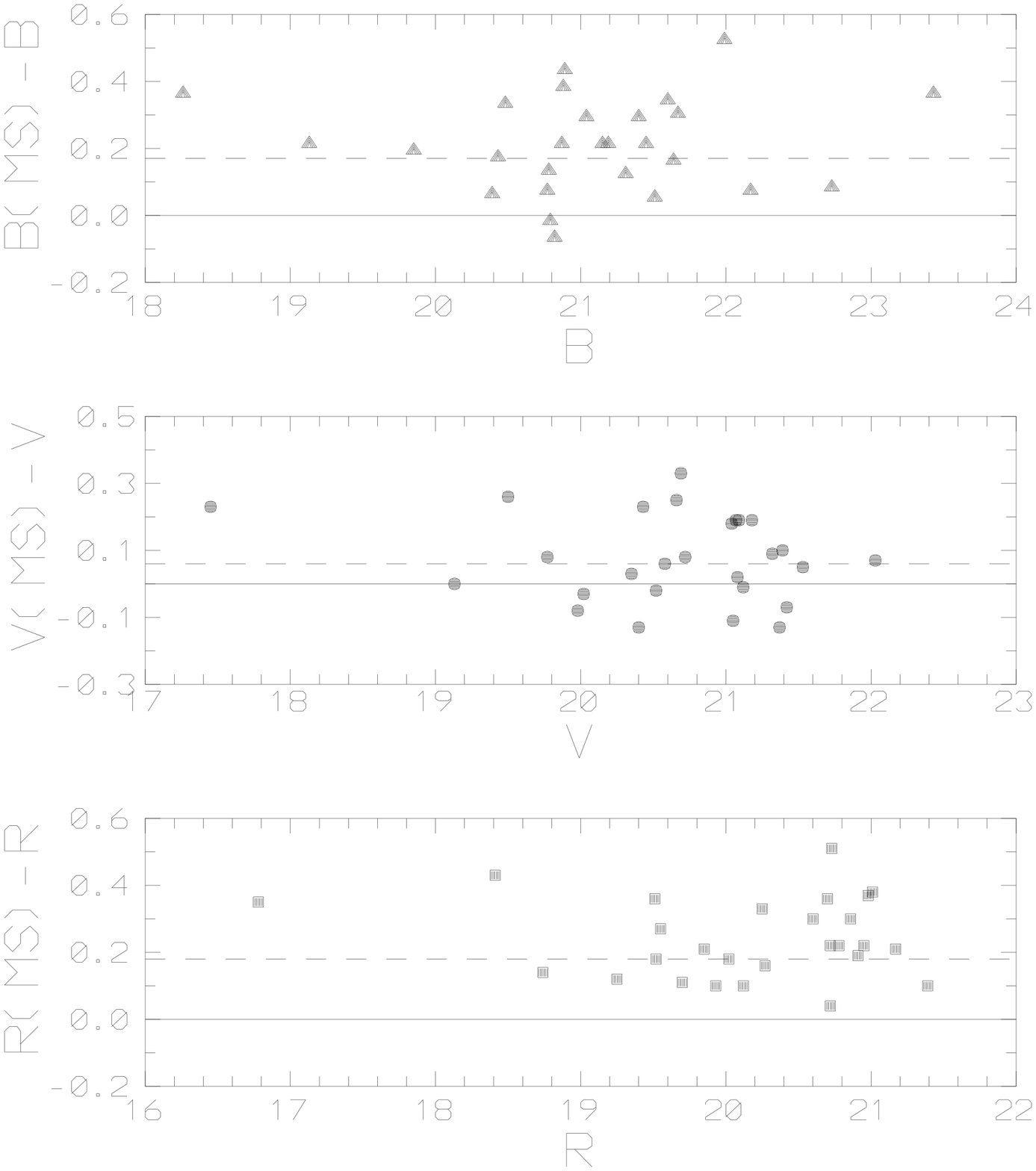}}
\caption{ 
Comparison of the magnitudes of the \cite{Sandage84q} stars 
measured in the present work with those obtained by \cite{Metcalfe91q}
in the  B, V and R bands.
}
\label{fig4}
\end{figure}

Using 27 isolated stars down to 23 mag from the \cite{Sandage84q} 
list, we compare the magnitudes of MS91 with our new photometry.
The results are shown in Fig\,\ref{fig4}. We derive the 
magnitude zero point shifts to be: 
\begin{eqnarray}
B(MS) - B(here) = 0.17 \pm 0.12 \\
V(MS) - V(here) = 0.06 \pm 0.14 \\
R(MS) - R(here) = 0.18 \pm 0.12
\end{eqnarray}

This also means that compared to the photometry of \cite{Sandage84q} 
our B magnitude zero point is fainter by 0.05\,mag, while our V magnitude zero 
point is the same.

\section{Color-magnitude diagrams}
Using an ellipse with a size of 3\arcmin\ by 5\arcmin\ and a position angle 
of $\sim 60\deg$  (as plotted in Fig.\,\ref{fig2}) we divided the 
observed field into two parts -- Ho\,IX and surroundings. The areas inside 
and outside the ellipse are about 12$\sq\arcmin$ and 18$\sq\arcmin$. We 
derived the 
B, V, and R magnitudes for 166 stars inside the ellipse and 60 stars 
outside it. The observed number of stars with V = 22-23\,mag inside the 
ellipse is about 3 time higher than outside it. In addition, we identified 
and measured 6 brighter stars as well as 11 round diffuse objects, which 
we suspect to be star cluster candidates. 
   
The data for the stars with V$<$22\,mag as well as for the diffuse 
objects are given in Table\,\ref{tab1}. 
The first column of Table\,\ref{tab1} 
contains the designation of the star. The letter S means that the number is 
from the \cite{Sandage84q} list. 
Numbers larger than S41 as well as those beginning by T correspond to 
stars noted in \cite{Georgiev91bq}. Other columns contain 
the X and Y coordinates in pixels, corresponding to the map in 
Fig.\,\ref{fig3}, V magnitudes and B-V and V-R color indices, and comments 
on the nature of the objects.  

\begin{figure}
\resizebox{\hsize}{!}{\includegraphics{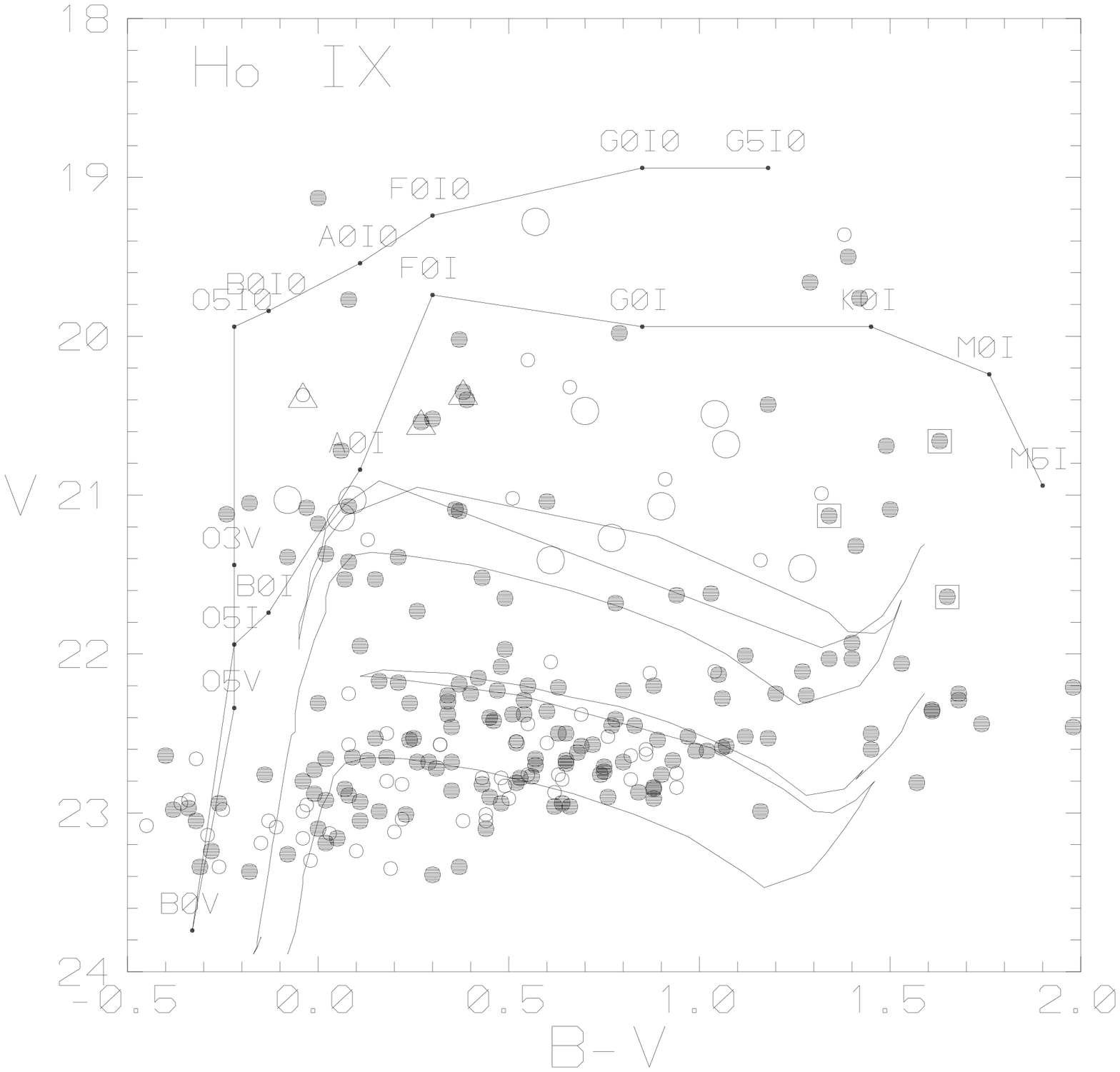}}
\resizebox{\hsize}{!}{\includegraphics{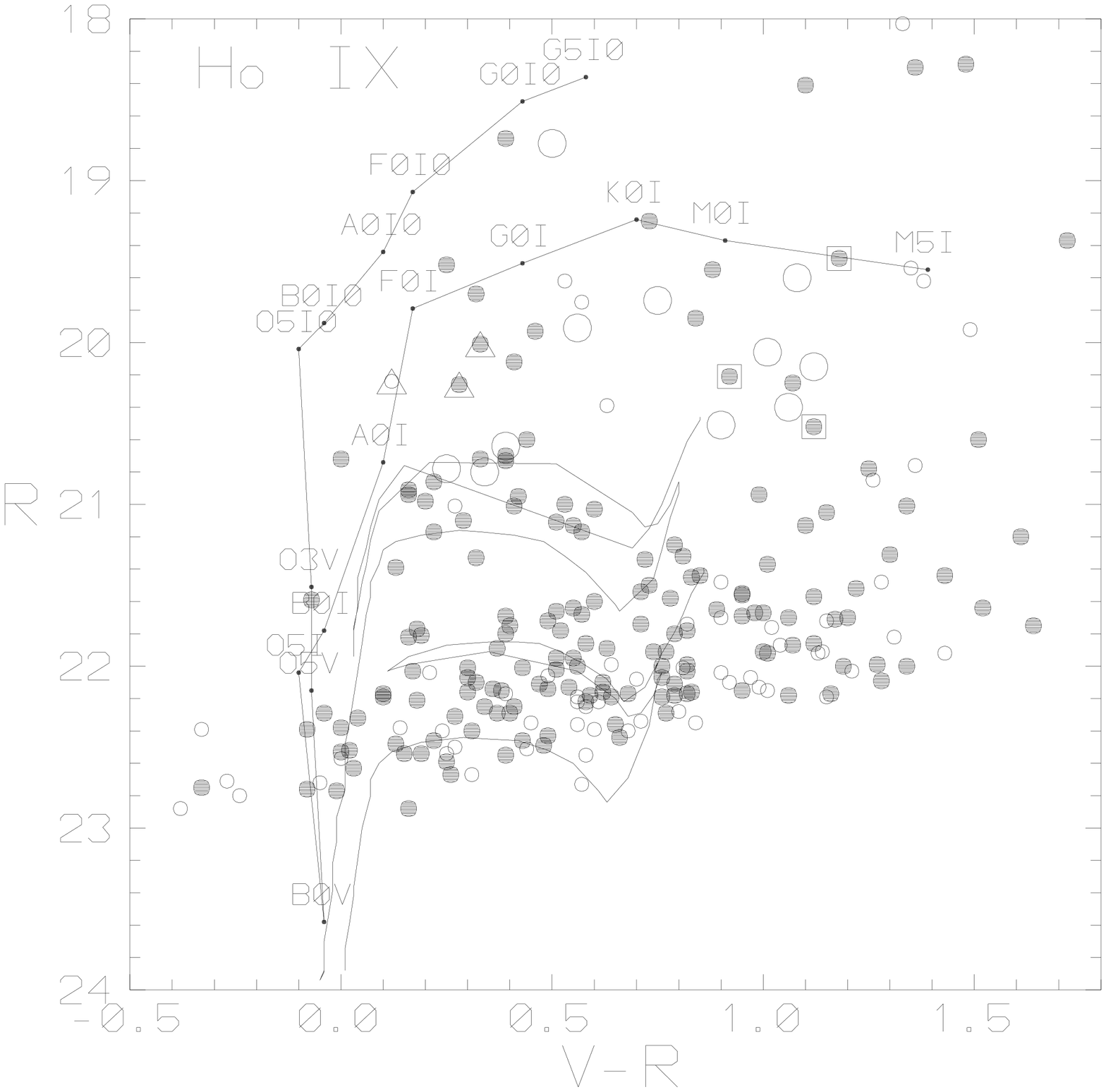}}
\caption{Photometric diagrams (B-V) vs. V (a, top) and (V-R) vs. R (b, bottom) 
of Ho\,IX. The star-like objects inside the ellipse in Fig.\,\ref{fig2} are 
represented by filled dots and those outside the ellipse by small circles. 
The diffuse objects are represented by large circles. The brightest blue 
and red supergiant candidates (see the text) are marked by additional 
triangles and squares, respectively. The lines represent the sequences of 
luminosity class I stars (z=0.02) taken from \cite{Schmidt-Kaler65q}  and 
\cite{Bertelli94q} isochrones for 20\,Myr and 40\,Myr (Z=0.004), shifted 
according to the reddening A$_B = 0.32$ and distance modulus 
$(M-m)_0 = 27.7$\,mag. 
}
\label{fig5}
\end{figure}

The color-magnitude diagrams of Ho\,IX are given in Fig.\,\ref{fig5}
and the (B-V) vs. (V-R) color-color diagram in Fig.\,\ref{fig6}. The 
lines represent the sequences of the brightest Milky Way stars from the 
data of \cite{Schmidt-Kaler65q}, and \cite{Straizys87q} (the 
Johnson-Morgan R magnitudes were transformed into the Kron-Cousins R system 
by the relations of \cite{Fernie83q}). Also plotted are isochrones for 20\,Myr 
and 40\,Myr \cite{Bertelli94}. The lines are shifted accounting for a 
foreground extinction A$_B = 0.32$ and a 
distance modulus $(M-m)_0 = 27.7$\,mag. 

The brightest diffuse objects are the globular cluster 
candidates in the periphery of M\,81, designated by \cite{Georgiev91aq} and 
\cite{Georgiev91bq} as C2, C4 and C5. The blue diffuse objects D3 and D7 
seem to be dense associations or young cluster candidates in Ho\,IX. 

Note that our diffuse object D1, placed to NE from the
bright star T1, coincides with the brightest X-ray source in the field 
of M\,81 \cite{Roberts00}. This object is visible also on the best B plate, 
taken with the 2m telescope of the Rozhen NAO and on the prints from the 
best B and V plates of the SAO 6\,m telescope \cite{Georgiev91b}, 
\cite{Georgiev92}. The object appears to be a distant face-on spiral galaxy 
of type Sc or Sd with apparent sizes of its "bulge" and very faint "disk" 
of diameters of $\sim 7\arcsec$ and $\sim 25\arcsec$, respectively. However, 
the inspection of the available material shows the presence of a blue spot, 
situated $\sim 2.2\arcsec$ west of the center of D2.  This blue spot
may be identified with the brightest X-ray source rather than the center
of D2.

The large bright red feature D4 is a H\,II region designated MH\,10 in 
the survey of \cite{Miller94q}.  The remaining objects D2, D5, D6, and D8 
may be distant galaxies, visible through Ho\,IX. If these diffuse 
objects are Ho\,IX members, the majority of them must have 
absolute magnitudes of about -6.5, which is consistent with compact 
star clusters at the distance modulus of Ho\,IX.

\section{On the distance to Ho\,IX}

In the absence of data for better distance estimators such as Cepheids or 
the tip of the red giant branch, we applied again the basic distance 
estimation method for star-forming galaxies, based 
on the apparent magnitudes of the brightest blue supergiant candidates 
(BBSC) and the brightest red supergiant candidates (BRSC) in Ho\,IX. 
This is a classic method of distance estimation \cite{Sandage74}, 
which has the advantage of requiring only a limited amount of observing 
time for a result, but is limited by calibration problems 
\cite{Karachentsev94} such as metallicity effects or the long known 
correlation of the brightness of the most luminous stars 
with the luminosity of their parent galaxies \cite{Humphreys83}.  
With a good empirical 
calibration the distance modulus $\mu = (m-M)_0 = 5 log D + 25$ can still 
be estimated to quite good accuracy. Here we use 
the recently published calibration of the method of the brightest stars 
(MBS), derived from published CCD surveys of nearby galaxies 
\cite{Georgiev97a}; \cite{Georgiev98}: 
\begin{eqnarray}
\mu = 1.39 (V3_r - A_v) - 0.39 (mbt-A_b) + 3.73 \pm 0.24 \\
\mu = 1.60 (B3_b - A_b) - 0.60 (mbt-A_b) + 3.83 \pm 0.41 \\
\mu = 1.30 (R3_r - A_r) - 0.30 (mbt-A_b) + 6.32 \pm 0.34 \\
\mu = 1.44 (R3_b - A_r) - 0.44 (mbt-A_b) + 4.69 \pm 0.51 
\end{eqnarray}

Here V3$_r$, B3$_b$, R3$_r$, and R3$_b$ are the mean apparent magnitudes of 
the 3 brightest blue supergiant candidates (BBSC) and the 3 brightest red 
supergiant candidates (BRSC) in the B,V, and R bands. 

For the extinction correction of the Ho\,IX data we use an  
extinction value of  $A_B = 0.32$ \cite{Paturel96}, which converts into 
$A_V = 0.76 A_B$, and $A_R = 0.57 A_B$ \cite{Fitzpatrick99}. 

One very critical part of this method is the selection of the 
brightest stars belonging to the target galaxy. One has to exclude 
foreground dwarfs, globular clusters of the target galaxy, and compact 
background galaxies.  Equally important is to select single stars and
not apparent double or multiple objects. Obviously, good photometric 
quality of the data and high spatial resolution is of essence here. 
For example: 
Some of the red stars around V = 19 - 20\,mag in the periphery of Ho\,IX were 
suspected to be the brightest red supergiants usable for distance 
estimation by \cite{Georgiev91bq}. Using our new data we checked
the magnitudes and colors of star \#40 of \cite{Sandage84q} and 
stars \#\# 44, 45, 46, 48, 49 of \cite{Georgiev91bq}. 
The B-V and V-R colors show that these stars are 
probable foreground red dwarfs. Therefore, the conclusion that the stars  
\#\# 34, 49 and 46 may be the {\it brightest red supergiant candidates} 
in Ho\,IX \cite{Georgiev91b} is wrong. Due to this misclassification, 
the distance modulus derived in \cite{Georgiev91bq} cannot be correct 
(see Table\,\ref{distances}).

\begin{figure}
\resizebox{\hsize}{!}{\includegraphics{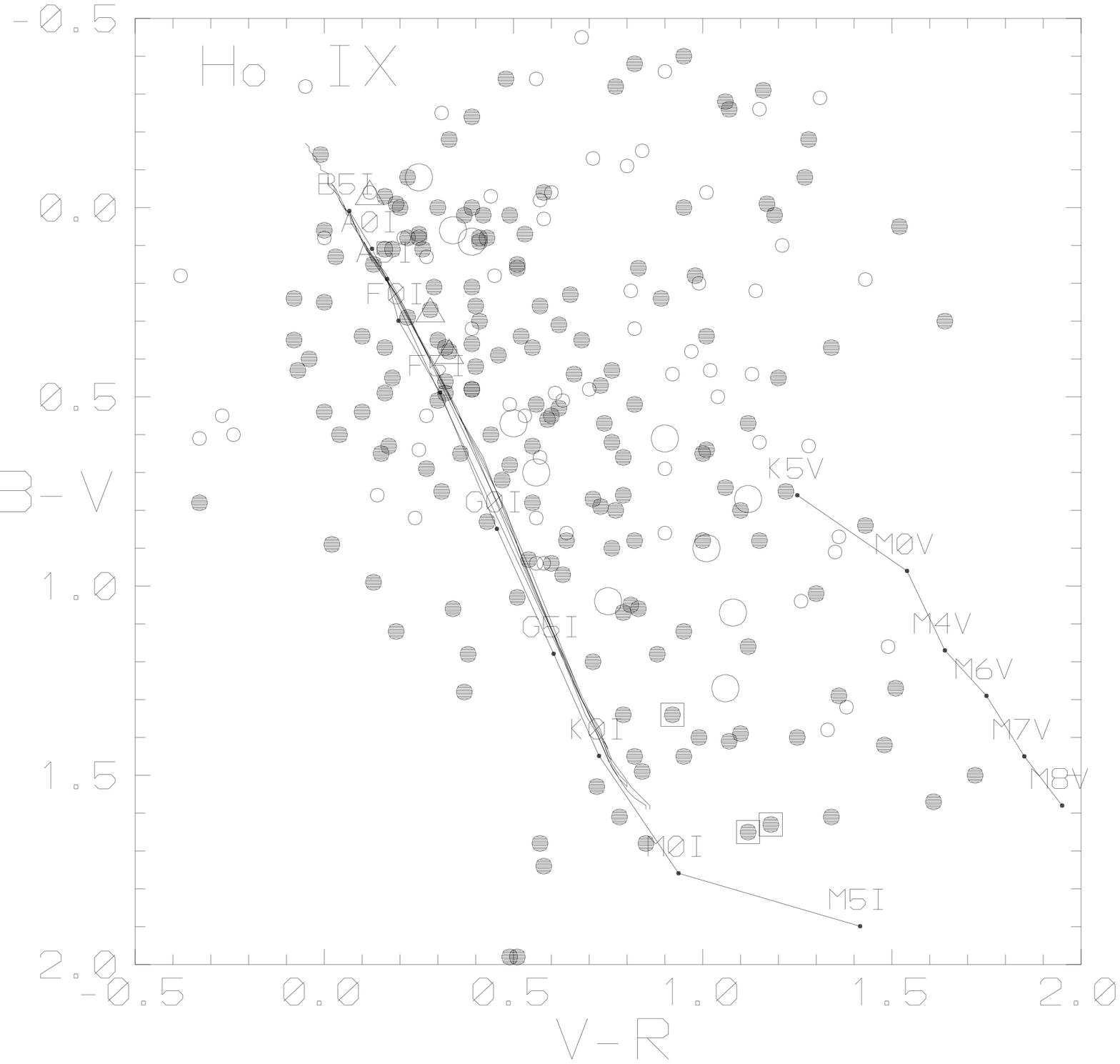}}
\caption{Color-color diagram of Ho\,IX.  The symbols are the same as in 
Fig.\,\ref{fig5}
}
\label{fig6}
\end{figure}

The use of the brightest red stars has advantages because these stars can 
be classified as BRSC and distinguished from galactic foreground stars 
using the color-color diagram (Fig.\,\ref{fig6}). 
For Ho\,IX the stars which are near to the line of the red supergiant stars 
of classes KI and MI in the color-color diagram are the stars S4, S9, and S41. 
From its colors the star S9 is a probable K supergiant, with a somewhat 
lower confidence level as S4 and S41, but still the next best candidate.
Another bright red star (\#138) must be 
classified as a red dwarf. The V3r and R3r magnitudes of the detached BRSC 
give $\mu = 27.58$ (D\,=\,3.28\,Mpc).
   
The brightest blue objects in the field of Ho\,IX are S16, S6, and S8. 
Analysing their shape, these stars appear slightly diffuse on our images.
This may indicate that these objects are not single stars but tight groups 
of several stars, similar to the tight groups observed in the LMC (e.g. 
\cite{Heydari-Malayeri94q}).
If we would use these stars as distance estimators we derive 
$\mu \sim 26.3$\,mag (D\,=\,1.8\,Mpc) which would place Ho\,IX near the 
fringes of the Local Group 
totally inconsistent with our result from the red supergiants. 
Based on their colors and profile shapes on our B,V, and R 
images we consider the stars S10, S36, and S50 as best candidates 
for the brightest BBSC. Using these stars, we derive a distance modulus of 
27.7\,mag. Such a distance modulus has generally lower weight than a distance 
modulus based on red supergiants due to the stronger clustering 
of the blue supergiants.
The tendency of blue supergiants to occur in tight groups, e.g.  
\cite{Heydari-Malayeri94}, is one of the basic disadvantages of using 
blue supergiants compared to the red supergiants.

Generally, using the stars S10, S36, and S50 as BBSC and the stars S4, 
S9, and S41 as BRSC and applying the four mentioned relation we estimate 
$\mu = 27.66 \pm 0.27$\,mag ($D = 3.40 \pm 0.6$\,mag). However, the 
uncertainty 
of the relations in respect to the different evolutionary status of the dwarf 
galaxies used in the calibration, as well as the uncertainty 
in the choice of the brightest stars, lower the accuracy of 
this method for deriving $\mu$ to about 0.4\,mag, corresponding to 
about 20\% in distance.

\begin{figure}
\resizebox{\hsize}{!}{\includegraphics{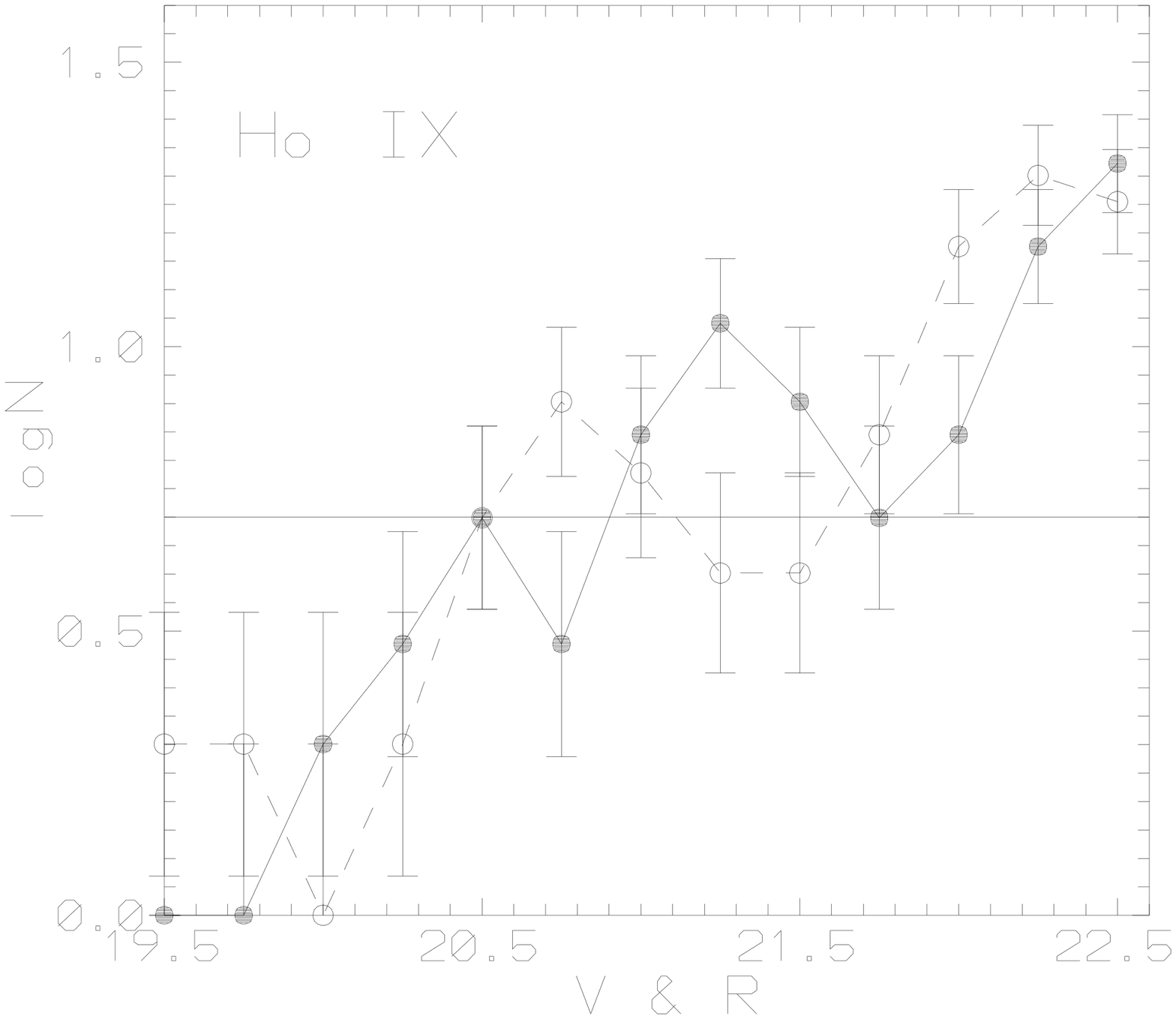}}
\caption{Differential luminosity functions of the blue stars with 
(B-V)$ < 0.5$ in the V band (solid line) and with (V-R) $ < 0.4$ in the 
R band (dashed line). The horizontal line corresponds to the distance 
estimator defined by five blue stars in a magnitude interval of 0.5 mag.  
}
\label{fig7}
\end{figure}

As was pointed out above, depending on the choice of the BBSC, the 
distance estimation may be changed by up to 100\%. The reason usually is 
that the brightest blue objects in the galaxies are young star clusters, 
compact associations or multiple stars seen as single ones. 
To avoid using multiple blue objects as distance indicators instead of 
single stars, we build the differential luminosity function of the bright 
blue stars (Fig.\,\ref{fig7}). Then we apply the magnitudes V(5b)$_0$ and 
R(5b)$_0$ for which the 
value of the smoothed luminosity functions represents five stars in an 
interval of 0.5\,mag. Using these magnitudes for the calibration of the 
luminosity function, the brightest blue stars as distance indicator of 
\cite{Georgiev98} are: 
\begin{eqnarray}
\mu = 1.85 V(5)_0 - 0.87 B(g)_0  + 1.08  \\
\mu = 1.68 R(5)_0 - 0.68 B(g)_0  + 2.21
\end{eqnarray}

{\bf
\begin{table}
\begin{tabular}{|lc|}
\hline
           & m$-$M \\
\hline
Sandage (1984) & 28.75  \\
Hopp \& Schulte-Ladbeck (1987) & $30.1 \pm 0.3$  \\
Sandage data recalibrated  & 29.4 \\
Davidge \& Jones (1989) & 27.5 \\
Georgiev et al. (1991)  & $27.67 \pm 0.25$\\
Karachentsev \& Tikhonov (1994) & 27.66 \\ 
this paper  &  $27.66 \pm 0.27$ \\
\hline
M\,81: Freedman et al. (1994) & $27.8 \pm 0.2$ \\
\hline
\end{tabular}
\caption{Short list of the distance estimates for Ho\,IX. The `Sandage data 
recalibrated' refers to a recalibration of Sandage's photometry by 
Hopp \& Schulte-Ladbeck (1987). The distance 
estimate of Davidge \& Jones (1989) is based on the luminosity function 
and not by applying the brightest supergiant method. For comparison 
a recent Cepheid distance to M\,81 is given, too. }
\label{distances}
\end{table}
}

In the present case using V(5b) $\sim$ 20.9\,mag and R(5b) $\sim$ 20.5\,mag 
we estimate m = 27.38\,mag and m = 26.76\,mag, corresponding to mean distance 
D = 2.63 Mpc. This is again a much lower distance than expected for a member 
of the M\,81 group. However, Ho\,IX is a small galaxy with 
ongoing star formation and its luminosity functions are therefore  
strongly distorted and shifted to bright magnitudes. 

Taking all these considerations into account the current best estimate 
of the distance of Ho\,IX is $\mu = 27.58$ (D\,=\,3.28\,Mpc).
This new distance estimation based on color selected red supergiants and 
a new calibration of the method is in agreement with \cite{Georgiev91bq}, 
obtained by other BBSC and BRSC, using the calibrations of 
\cite{Karachentsev94q}.  The distance is also consistent with 
membership of the M\,81 galaxy group as suggested by the proximity to 
M\,81 and the possible correspondence to the HI cloud \cite{Yun94}.

It is important to note here that difference in the photometric 
zeropoints, differences in the choice of stars for the BBSC and BRSC, 
and different empirical calibrations of the relations can cancel 
out to some degree.  We compiled the distance estimate of Ho\,IX in 
Table\,\ref{distances}, clearly shows this point. We showed that 
the photometry of \cite{Sandage84q} is essentially correct, as 
already suspected by \cite{Metcalfe91q}. Still, the distance modulus 
is very large, which is an effect of the selection of stars and the 
calibration of the brightest supergiant method used in that paper.
As another example \cite{Georgiev91bq} 
and the present paper give essentially the same distance modulus, but the 
earlier result is based on a wrong selection of stars and a different 
calibration of the method.

Note, that HST has not yet improved the distance estimate of Ho\,IX. 
The recent 
HST WFPC2 (\cite{Karachentsev02}; \cite{Makarova02}) photometry of 
Ho\,IX did not yield a distance, since the red 
giant branch was not detected and the supergiant criterion could not 
be applied due to the incomplete coverage of the galaxy.

\section{Recent star formation history of Ho\,IX}
The properties of our CMD and that of other published CMDs of Ho\,IX 
by \cite{Hopp87q} and \cite{Davidge89q} are very similar. Ho\,IX is 
far less active than classical star forming or starbursting dwarf 
galaxies, e.g. \cite{Vallenari96q} or \cite{Hopp95q}.
   
The region of increased concentration of blue stars at V$\sim 23$\,mag and 
B-V $\sim -0.3$ as well as R $\sim 22.5$\,mag and V-R $\sim 0$  
the V vs. (V-B) and R vs. (V-R) color-magnitude diagrams corresponds well  
to the main sequence location of B0V-O5V stars having absolute 
magnitudes of about $-4.5$ to $-5$\,mag. 
The isochrones of \cite{Bertelli94q} are superimposed on the CMDs. 
The turn-off points of the evolved stellar population indicate an age 
of about 40\,Myr. 

The CMD is unfortunately not deep enough to draw more detailed conclusions 
on the star formation history directly.  Together with the results from 
the literature some interesting points can nevertheless be noted.  
The surface brightness of Ho\,IX is quite low, putting it well into 
the regime of low surface brightness galaxies, still the intermediate 
age population seems to be similar to that of the LMC \cite{Davidge89}, 
a much more massive and higher surface brightness object. If true, this 
implies a significant star formation event a few Gyr ago, like the 
possibly tidally triggered burst of star formation in the LMC about 2\,Gyr 
ago (e.g. \cite{Bomans95r}).   
The patchy structure of Ho\,IX and the location of the most significant 
region of relatively recent star formation may hint at tidal influence 
of M\,81, but this remains speculation until much deeper imaging data 
reveal a much more detailed picture of the star formation history of 
Ho\,IX.

\section{Summary and conclusions}
In this paper we investigated the widely different photometric zeropoints 
of several photometries of the resolved stellar content of the dwarf 
galaxy Ho\,IX.  It turned out that the zeropoints of the photometric 
work of \cite{Sandage84q} is consistent with our CCD based photometry.  
Due to its immediate proximity to M\,81 on the sky, Ho\,IX is a prime 
candidate for investigating tidal effects on the evolution of dwarf galaxies, 
as long as Ho\,IX is not offset significantly in front of or behind M\,81.  

The coincidence of Ho\,IX with a HI cloud at the `right' velocity is 
not a fully convincing argument due to the low level of recent star formation 
activity in Ho\,IX.  Ho\,IX could have no significant HI left 
and the detected HI would be unrelated to Ho\,IX, belonging to one of the 
tidal debris HI clouds seen around M\,81, M\,82, and NGC\,3077.

In the absence of other distance indicators we used our photometry and 
a new calibration of the brightest red and blue supergiant criterion 
to estimate the distance to Ho\,IX.  
The brightest blue supergiant method (using CMD and luminosity function) 
yielded low distances, inconsistent with a membership of Ho\,IX of the 
M\,81 group.  Due to the low recent star formation rate and extended 
nature of several blue supergiant candidates, we regard this result 
as unreliable.

We improved the selection of the brightest red supergiants by using the 
B-V versus V-R color-color diagram (selection against galactic foreground 
stars), and again used the new calibration of \cite{Georgiev98q}.  
The resulting distance of $\mu = 27.66 \pm 0.27$\,mag ($D = 3.40 \pm 0.6$\,Mpc)
would place Ho\,IX in the immediate vicinity of M\,81.  

Still, the error in this distance is too large to declare the matter 
settled.  Photometry which makes it possible to determine the distance from 
the tip of the red giant branch is highly needed.  Such high quality data 
would also make it possible to search for tidally induced bursts in the 
star formation history of Ho\,IX.

Using HST WFPC2 snapshot observation a very good quality CMD has 
been produced (\cite{Karachentsev02}; \cite{Makarova02}). 
Despite being much deeper than our observations, no red giant 
branch was detected and no independent distance estimate for Ho\,IX could 
me made. Still, the CMD seems to be consistent with a distance close to that of
M\,81. The absence of a giant branch in the CMD remains puzzling. 
Clearly, the star formation history and therefore 
the nature of Ho\,IX is far from understood.

\begin{acknowledgements}
TSG thanks the Graduate Schools GRK\,118 `The Magellanic System, Galaxy 
interaction, and the Evolution of Dwarf Galaxies' and GRK\,787 
`Galaxy Groups as Laboratories for baryonic and Dark Matter' 
for support 
during his visits to Bochum, and the staff of the Rozhen MAO for their help
during the observations. The authors thank R.-J. Dettmar 
for critical reading of the manuscript and several helpful comments.  

The Digitized Sky Surveys were produced at the Space Telescope Science 
Institute under U.S. Government grant NAG W-2166. The images of these surveys 
are based on photographic data obtained using the Oschin Schmidt Telescope on 
Palomar Mountain and the UK Schmidt Telescope. 
The Second Palomar Observatory Sky Survey (POSS-II) was made by the California 
Institute of Technology with funds from the National Science Foundation, the 
National Geographic Society, the Sloan Foundation, the Samuel Oschin 
Foundation, and the Eastman Kodak Corporation. 

This research has made use of the NASA/IPAC Extragalactic Database (NED) 
which is operated by the Jet Propulsion Laboratory, California Institute of 
Technology, under contract with the National Aeronautics and Space 
Administration.
\end{acknowledgements}

\appendix
\section{Notes on the methods}
The aperture stellar photometry in this paper is made simultaneously in the 
three rebinned frames (in B, V and R bands) by the program UMAG (Upper 
MAGnitudes). This software is especially developed for photometry in 
crowded fields of galaxies where the number of star-like objects may be 
comparable to or lower than the number of slightly diffuse ones. 
Under these conditions PSF photometry is not as efficient as usual  
(\cite{Georgiev95br}; \cite{Notni96r}; \cite{Georgiev97br}). 

The photometry method realized in the program UMAG is a 
generalization of the aperture photometry method. A fast algorithm for 
constructing the magnitude growing curve (MGC) of the image within the 
boundaries of a circular photometric aperture is implemented, based on 
histogram transforming. UMAG gives two kind of results -- instrumental 
magnitudes and shape parameters, as well as MGCs of the images. 
   
The program UMAG builds a special kind of MGCs. Each value of the 
usual MGC is derived using only one background plane, which corresponds 
to the periphery pixels of the photometric aperture. For this reason all 
values of the usual MGC have equal shifts due to the error of the 
background determination. Each value of our "upper" MGC is an integral 
of the image within a circular sub-aperture (inside the photometric 
aperture) and above the background plane. This background is determined on the 
periphery of the sub-aperture. So each value of the upper MGC depends on 
its "inner" background (and its error). The values of the upper MGC may 
be called "upper" magnitudes, because each of them is determined using 
a background plane which is above the background around the image. 

The last points of both kinds of MGC coincide. A  
photometric method which uses these values as instrumental magnitudes is the 
usual aperture photometry. However, the upper MGC is less steep than the 
normal MGC and its values depend more strongly on the shape of the image. 
Therefore, it provides information for image classification. We use 
the first and second derivatives of the upper MGC and distribute the 
images into shape classes -- very sharp (stellar images which may be 
corrupted by cosmics), star like, cluster like, irregular (possible 
unresolved stellar associations or background galaxies). 

Star images that are similar in their MGCs (normal 
MGC, as well as the upper MGC) are similar (proportional) in shape. However, 
the pixel sampling of the central parts of the stellar images disturbs 
the similarity and conserves some proportionality in a narrow magnitude 
interval, usually 6-7 magnitudes. For this reason we use small 
aperture diameters (no less than $2 \times$ FWHM of the PSF) for an interval 
of about 6 magnitudes above the limiting magnitude of the frame. 
Brighter objects must be measured with a larger aperture. 
   
The program UMAG determines the instrumental magnitudes by averaging 
the values of 6\% of the brightest tails of the MGCs and adding an 
aperture correction. The mean values of the aperture correction in each 
band is calculated by the program. It uses isolated star images for which 
the instrumental magnitudes are determined and then extrapolated  
to a magnitude determined with an aperture with infinite radius. 
The list of 
isolated stars and their full instrumental magnitude, as well as 
the constants of the color transform must be prepared as input files 
for the final photometry.
Then the program UMAG gives directly the standard magnitudes and colors 
of the images (as well as their shape parameters and upper MGC). 
   
The difference between UMAG photometry and e.g. DAOPHOT aperture 
photometry \cite{Stetson87} is only in the number of the background 
pixels. DAOPHOT uses rings of pixels around the photometric aperture while 
UMAG uses just the peripheral pixels of the same aperture and does not 
need additional background space. A similar method of photometry was 
applied independently by \cite{Notni96q} in the crowded field of 
the galaxy M\,82. 
A comparison of the UMAG and DAOPHOT PSF photometry of stellar images 
in the crowded fields of 5 resolved dwarf galaxies is given by 
\cite{Makarova97q}. It shows a good agreement in the range 18-24 mag within a 
standard error of 0.1 mag. The results from both programs differ 
primarily when the objects are slightly diffuse. UMAG underestimates and 
PSF DAOPHOT overestimates their brightness. 



\end{document}